# Coupled solitons of intense high-frequency and low-frequency waves in Zakharov-type systems

**Gromov Evgeny**[1, a)] **and Malomed Boris**[2,3]

[1]National Research University Higher School of Economics (HSE), Nizhny Novgorod 603155, Russia

[2]Department of Physical Electronics, School of Electrical Engineering, Faculty of Engineering, Tel Aviv University, Tel Aviv 69978, Israel

[3]Laboratory of Nonlinear-Optical Informatics, ITMO University, St. Petersburg 197101, Russia

**Abstract**

One-parameter families of exact two-component solitary-wave solutions for interacting high-frequency (HF) and low-frequency (LF) waves are found in the framework of Zakharov-type models, which couple the nonlinear Schrödinger equation (NLSE) for intense HF waves to the Boussinesq (Bq) or Korteweg - de Vries (KdV) equation for the LF component through quadratic terms. The systems apply, in particular, to the interaction of surface (HF) and internal (LF) waves in stratified fluids. These solutions are two-component generalizations of the single-component Bq and KdV solitons. Perturbed dynamics and stability of the solitary waves are studied in detail by means of analytical and numerical methods. Essentially, they are stable against separation of the HF and LF components if the latter one is shaped as a potential well acting on the HF field, and unstable, against splitting of the two components, with a barrier-shaped LF one. Collisions between the solitary waves are studied by means of direct simulations, demonstrating a trend to merger of in-phase solitons, and elastic interactions of out-of-phase ones.

**Among the great variety of models which give rise to solitary waves (we call them "solitons", as it is widely accepted in physics literature, even if the corresponding models are not integrable), an important class is based on coupled equations for the interaction of high- and low-frequency (HF and LF) waves. Commonly called Zakharov systems, referring to the pioneering paper by V. E. Zakharov (1971), which had introduced this system as a fundamental model for the interaction of Langmuir (HF) and ion-acoustic (LF) waves in plasmas, these models find a large number of other physical realizations, such as the coupling of vibrational (HF) and longitudinal acoustic (LF) perturbations in long molecules (closely related to the famous Davydov's model of the DNA dynamics), and the interaction of HF surface waves and LF internal waves in stratified fluids, with important applications to oceanography. In the latter context, as well as in other physically relevant settings, the Zakharov's system is generalized by adding intrinsic HF and LH nonlinearities to the nonlinear HF-LF coupling. Further, the interplay of the nonlinearities with the inner dispersions of the HF and LF components may naturally create solitons, which are fundamentally important excitations in the underlying physical systems. Previously, particular soliton solutions of Zakharov's systems were reported, including the well-known**

---

[a)] Author to whom correspondence should be addressed. Electronic mail: egromov@hse.ru

**Davydov-Scott soliton. However, they were found under quite restrictive special conditions, such as equal strengths of the dispersion and nonlinearity in the LF component. In this work, we aim to produce new families of two-component HF-LF solitons, found in an exact analytical form under more general conditions, although these solution families are not exhaustive, as still more general ones may exist, being accessible solely a numerical form. This is done for two relevant forms of the Zakharov system, *viz.*, ones including the nonlinear Schrödinger equation for the HF component, coupled to the bidirectional Boussinesq (Bq) or unidirectional Korteweg – de Vries (KdV) equation for the LF waves. Further, for the systems including the Bq and KdV equations, we perform a detailed study of the solitons' stability, concluding that the family splits into subfamilies which are stable or not against spatial separation of the HF and LF components. We also study, by means of direct simulations, collisions between stable solitons, which is a relevant aspect of the soliton dynamics in non-integrable systems, such as those addressed in the present work.**

## I. Introduction

The common interest in solitons (stable nonlinear localized waves) is caused, among other reasons, by their remarkable ability to travel large distances keeping a robust form. Solitons arise in models of the wave propagation in diverse dispersive nonlinear media, including surface waves (SWs) on deep and shallow water, internal waves (IWs) in stratified liquids, Langmuir and ion-sound waves in plasma, pulse and beams in nonlinear photonics, matter waves in Bose-Einstein condensates, electromagnetic waves in long Josephson junctions, spin waves in magnetically ordered materials, etc [1-8]. In rigorous mathematical analysis, term "solitons" is usually reserved only for exact solutions of integrable models [1], but in physics literature solitary waves in nonintegrable systems are also called solitons. Following the latter convention, we use this term below.

In the second-order approximation of the nonlinear dispersion wave theory [8], dynamics of intense high-frequency (HF) waves, such as SWs on deep water, Langmuir waves in plasma, high-power optical signals in fibers and planar waveguides, etc., is based on the nonlinear Schrodinger equation (NLSE) [9,10]. In the framework of this equation, HF envelope solitons arise from the balance of the linear second-order dispersion and the cubic self-focusing.

Dynamics of intense low-frequency (LF) waves, such as IWs in stratified liquids, SWs on shallow water and ion-sound waves in plasma, is modeled by the bidirectional Boussinesq (Bq) equation or, as a limit case, the unidirectional Korteweg - de Vries (KdV) equation. Soliton solutions of these equations originate from the balance of nonlinearity and dispersion of the LF waves.

In many physically relevant settings, intense HF waves are naturally coupled to the propagation of LF waves. This is possible because, while the frequencies and wavenumbers of the two wave modes of the two types are widely different, their phase and group velocities may happen to be close, thus providing resonant enhancement of the coupling. A generic class of models of the HF-LF interaction is provided by systems of the Zakharov's type, which include the linear Schrödinger equation coupled to the Bq/KdV equation [11]. In these systems, the HF field is governed by the Schrödinger equation, with the coupling represented by the potential term produced by the LF waves [12-14]. For HF Langmuir waves in plasmas, the potential term accounts for the variation of the plasma density caused by LF ion-sound waves, while for SWs in stratified liquids the potential is induced by the flow current on the surface of the liquid caused by the IWs. In these systems, the LF component is governed by the bidirectional Boussinesq (Bq) equation or unidirectional KdV equation (the derivations can be found in Refs. [15-18] and [19], respectively). In addition to the proper nonlinearity and dispersion of the LF waves, these equations include a quadratic source term induced by the HF waves, which accounts for the coupling to them.

Soliton solutions of these systems were obtained in the following cases: (i) for the relation of the nonlinearity and dispersion coefficients of the LF wave ($\beta$ and $\gamma$, respectively, see section 2 below) $\beta/\gamma = 1$, the two-component (vector) Davydov-Scott (DS) soliton solution was found [20]; (ii) for $\beta/\gamma \neq 1$ and a small amplitude of the HF field, two-component asymptotic multihump solitons were found [14]; (iii) neglecting the strictional action of the HF components on the LF waves, which induce a potential well in the HF equation, localized linear stationary states of the HF field were found too [19].

On the other hand, the description of intense SWs on deep water must take into account the self-phase-modulation term [19], which replaces the linear Schrodinger equation for the HF field in the Zakharov system by the NLSE. A similar generalization of the Zakharov system for SWs in elastic solids was previously introduced in Ref. 21.

In this work the interaction of intense HF and LF waves is considered in the framework of two modified Zakharov systems of coupled NLS-Bq/KdV equations, which are introduced in Sections II and III, respectively. New families of exact analytical two-component soliton solutions, each with one free parameter, are found in Sections II.A and III.A, for an arbitrary relative strength of the nonlinearity and dispersion of the LF waves. These solutions are two-component generalizations of the single-component soliton solutions of the Bq/KdV equations. They are most generic solutions which can be produced analytically, although still more general ones may plausibly be found in a numerical form, with two free parameters, representing the amplitude and velocity of the solitons (such a numerical problem is not considered here). In the framework of the coupled NLS-Bq/KdV system, perturbed dynamics of the two-component solitons, induced by a spatial shift of their HF component against the LF one (this is a critical perturbation which may cause instability of the solitons), and by a change of the latter's amplitude, is considered in Sections II.B and III.B, by means of a perturbative approximation. Section IV reports numerical results, which corroborate the analytical predictions concerning the soliton stability. Collisions between stable solitons are studied too in Section IV, by means of direct simulations. The paper is concluded by Section V.

## II. The system of coupled NLS-Bq equations

We start by considering the unidirectional copropagation of a slowly varying envelope, $U(x,t)$, of the complex HF wave field, $U(x,t)\exp(ik_0 x - i\omega_0 t)$, and its real LF counterpart, $n(x,t)$ (effectively, it is a local perturbation of the refractive index in the HF equation). If the HF and LF fields represent the SW and IW, respectively in the ocean, the corresponding Zakharov system, under the above-mentioned condition of the group-velocity resonance, is composed of the NLSE for the SW and Bq equation for the IW, coupled by the usual (quadratic) terms [17,18]:

$$2i\left(\frac{\partial U}{\partial t} + V_{HF}\frac{\partial U}{\partial x}\right) - \frac{\partial^2 U}{\partial x^2} - 2\alpha|U|^2 U + 2nU = 0, \tag{1a}$$

$$\frac{\partial^2 n}{\partial t^2} - V_{LF}^2 \frac{\partial^2 n}{\partial x^2} + 6\beta \frac{\partial^2 (n^2)}{\partial x^2} - \gamma \frac{\partial^4 n}{\partial x^4} = \varepsilon \frac{\partial^2 (|U|^2)}{\partial x^2}, \quad (1b)$$

where $V_{HF} = (\partial \omega_{HF} / \partial k)_{k_0}$ is the HF group velocity, $V_{LF} = \omega_{LF} / k_{LF}$ is the LF phase velocity, $\alpha$ is the coefficient of the cubic nonlinearity (self-phase modulation) of the HF waves, $\beta$ is the nonlinearity coefficient of the LF waves, $\gamma$ is the dispersion of the LF waves, and $\varepsilon$ is the HF striction coefficient.

Below, we will use the representation of complex equation (1a) in the form of two real equations, obtained by means of the where the complex field is represented in the Madelung's substitution, $U \equiv |U| \exp(i\varphi)$, with $K \equiv \partial \varphi / \partial x$ being the respective wavenumber:

$$\frac{\partial |U|}{\partial t} + V_{HF} \frac{\partial |U|}{\partial x} - \frac{1}{2} |U| \frac{\partial K}{\partial x} - K \frac{\partial |U|}{\partial x} = 0, \quad (1c)$$

$$\frac{\partial K}{\partial t} + V_{HF} \frac{\partial K}{\partial x} + \frac{1}{2} \frac{\partial}{\partial x} \left( \frac{1}{|U|} \frac{\partial^2 |U|}{\partial x^2} \right) - K \frac{\partial K}{\partial x} + 2\alpha |U| \frac{\partial |U|}{\partial x} - \frac{\partial n}{\partial x} = 0 \quad (1d)$$

(the latter equation is derived by means of additional differentiation in $x$ and using identity $\partial / \partial x (\partial \varphi / \partial t) \equiv \partial K / \partial t$).

For $\varepsilon = 0$ and without the LF component, $n = 0$, system (1) is reduced to the NLSE which gives rise to the fundamental soliton solution:

$$U = \sqrt{\lambda / \alpha} \operatorname{sech}\left[\sqrt{\lambda}(x - Vt)\right] \exp\left[-i(1/2)(\lambda + V_{HF}^2 - V^2)t + i(V_{HF} - V)x\right]. \quad (2)$$

In another obvious limit, without the HF component, $U = 0$, system (1) is reduced to the Bq equation with the respective soliton:

$$n = -\frac{\gamma}{\beta \Delta^2} \operatorname{sech}^2\left(\frac{x - Vt}{\Delta}\right), \quad V^2 = V_{LF}^2 + \frac{\gamma}{\Delta^2}, \quad \Delta = 2\sqrt{\lambda / (V^2 - V_{LF}^2)}. \quad (3)$$

When both fields $U$ and $n$ are present, the following soliton solutions of Eq. (1) were analyzed, neglecting of self-phase modulation of the HF waves ($\alpha = 0$):

(i) for $\gamma / \beta = 1$, the two-component Davydov-Scott soliton solution with two free parameters $\lambda, A_0$ was found [20]:

$$A = A_0 \operatorname{sech}\left[\sqrt{\lambda}(x - Vt)\right] \exp\left[-i(1/2)(\lambda + V_{HF}^2 - V^2)t + i(V_{HF} - V)x\right],$$
$$n = -\lambda \operatorname{sech}^2\left[\sqrt{\lambda}(x - Vt)\right] < 0, \qquad V = \pm\sqrt{V_{LF}^2 + 4\gamma\lambda - \varepsilon A_0^2 / \lambda}; \quad (4)$$

(ii) for $\gamma / \beta \neq 1$ and small amplitude of the HF field, $|U|/n \ll 1$, asymptotic multihump solitons were found in Ref. 14;

(iii) for $\varepsilon = 0$ and an LF solution in form of a hole (alias potential well, with $n < 0$), localized stationary states of the linear HF equation were found in Ref. 17.

In the next section, a family of new exact two-component soliton solutions is found in the framework of the modified Zakharov system (1) with any ratio of the nonlinearity and dispersion coefficients of the LF component ($\beta$ and $\gamma$).

## A. Coupled-soliton solutions of the NLS-Bq system

In the reference system moving with velocity $V$, i.e., with coordinate $\xi = x - Vt$, a solution to Eq. (1) is looked for with the HF component in the respective stationary form,

$$U = \psi(\xi)\exp[-i\Omega t - i(V - V_{HF})\xi], \qquad (5)$$

which leads to a s system of ordinary differential equations:

$$\frac{d^2\psi}{d\xi^2} - \lambda\psi + 2\alpha\psi^3 - 2n\psi = 0, \qquad (6a)$$

$$\frac{d}{d\xi}\left[\gamma\frac{d^2n}{d\xi^2} - (V^2 - V_{LF}^2)n - 6\beta n^2 + \varepsilon\psi^2\right] = 0, \qquad (6b)$$

where $\lambda \equiv 2\Omega - (V - V_{HF})^2$ is a free parameter. Integrating Eq. (6b) under the localization condition, $(n,\psi)|_{\xi \to -\infty} \to 0$, and defining $\eta = \xi/\Delta$, system (6) reduces to

$$\frac{1}{\Delta^2}\frac{d^2\psi}{d\eta^2} - \lambda\psi + 2\alpha\psi^3 - 2n\psi = 0, \qquad (7a)$$

$$\frac{\gamma}{\Delta^2}\frac{d^2n}{d\eta^2} - (V^2 - V_{LF}^2)n - 6\beta n^2 + \varepsilon\psi^2 = 0. \qquad (7b)$$

A solution of system (7) can be looked for as

$$\psi = A\,\text{sech}\,\eta, \quad n = B\,\text{sech}^2\eta. \qquad (8)$$

Substituting Eq. (8) in (7), one arrives at a system of algebraic equations:

$$\alpha A^2 = \lambda\left(1 - \frac{\gamma}{\beta}\right), \qquad (9)$$

$$\frac{1}{\Delta^2} = \lambda > 0, \qquad (10)$$

$$B = -\frac{\lambda\gamma}{\beta}, \qquad (11)$$

$$V^2 = V_{LF}^2 - 4\beta B + \frac{\varepsilon A^2}{B}. \qquad (12)$$

For $\alpha = 0$, Eq. (9) amounts to $\gamma/\beta = 1$, leaving the amplitude of the HF component, $A$, arbitrary. This solution corresponds to the above-mentioned Davydov-Scott soliton (4) with two free parameters, $\lambda$ and $A$ [20].

For $\alpha \neq 0$ Eq. (9) implies sign condition $(1/\alpha)(1 - \gamma/\beta) \geq 0$, the amplitude of the HF component being $A = \pm\sqrt{(\lambda/\alpha)(1 - \gamma/\beta)}$. This gives rise to the new exact two-component solitons for coupled HF and LF waves with one free parameter, $\lambda > 0$:

$$U = \pm\sqrt{\frac{\lambda}{\alpha}\left(1 - \frac{\gamma}{\beta}\right)}\,\text{sech}\left[\sqrt{\lambda}(x - Vt)\right]\exp\left[-\frac{i}{2}(\lambda + V_{HF}^2 - V^2)t + i(V_{HF} - V)x\right],$$

$$n = -\frac{\lambda\gamma}{\beta}\operatorname{sech}^2\left[\sqrt{\lambda}(x-Vt)\right], \quad V = \pm\sqrt{V_{\text{LF}}^2 + 4\gamma\lambda - \frac{\varepsilon}{\alpha}\left(\frac{\beta}{\gamma}-1\right)}. \tag{13}$$

At $\gamma/\beta = 1$, the HF component of the soliton vanishes, $U = 0$, reducing soliton (13) to the LF soliton (3) of the Bq equation, therefore solution (13) may be interpreted as a vector generalization of scalar solution (3). For $\gamma/\beta < 1$, soliton (13) exists under the condition that the cubic nonlinearity of the HF field is self-focusing, $\alpha > 0$, while for $\gamma/\beta > 1$ the cubic nonlinearity must be defocusing, $\alpha < 0$. For weak dispersion and nonlinearity of the LF waves, and weak striction force created by the HF field in the Bq equation for the LF component, the velocity of soliton (13) is close to the velocity of the linear LF waves, $V \approx V_{\text{LF}} + 2\gamma\lambda/V_{\text{LF}} - (\varepsilon/2V_{\text{LF}}\alpha)(\beta/\gamma - 1)$.

## B. Dynamics of perturbed solitons with a relative shift between the HF and LF components of the NLS-Bq system

Straightforward application of the spatial integration to Eqs. (1c,1d,1b), written in the Madelung form, with zero boundary conditions at infinity, $(n,U)|_{x\to\pm\infty} \to 0$, gives rise to the following exact integral relations for field moments:

$$\frac{dN}{dt} \equiv \frac{d}{dt}\int_{-\infty}^{+\infty}|U|^2 dx = 0, \quad \frac{d^2}{dt^2}\int_{-\infty}^{+\infty} n\, dx = 0, \tag{14}$$

$$\frac{d}{dt}\int_{-\infty}^{+\infty} K|U|^2 dx = -\int_{-\infty}^{+\infty} n\frac{\partial(|U|^2)}{\partial x}dx, \tag{15}$$

$$\frac{d}{dt}\int_{-\infty}^{+\infty} x|U|^2 dx = -\int_{-\infty}^{+\infty}(V_{\text{HF}} + K)|U|^2 dx, \tag{16}$$

The dynamical equations for moments were used in many works as a basis for the description of perturbed evolution of solitons in the adiabatic approximation [22].

In the present situation, solutions for a perturbed compound soliton of the NLS-Bq system may be looked for in the form of sech-like pulses, with a shift of the HF component relative to the LF counterpart, keeping the fixed width:

$$U = A\operatorname{sech}\left[\sqrt{\lambda}(x - Vt - \bar{x}(t))\right]\exp\left[-i(1/2)(\lambda + V_{\text{HF}}^2 - V^2)t + i(V_{\text{HF}} - V)x + ik(t)x\right],$$

$$n = B\operatorname{sech}^2\left[\sqrt{\lambda}(x - Vt)\right], \tag{17}$$

where $\bar{x}(t) = N^{-1}\int_{-\infty}^{+\infty} x|U|^2 dx$ is the center-of-mass coordinate of the HF component, $k(t) \equiv K + V$, $A \equiv \pm\sqrt{(\lambda/\alpha)(1 - \gamma/\beta)}$, $B \equiv -\lambda\gamma/\beta$, $V^2 \equiv V_{\text{LF}}^2 - 4\beta B + \varepsilon A^2/B$. The shift between the two components, $\bar{x}(t)$, along with its dynamically-conjugate momentum, $k(t)$, is a crucially important ingredient of the perturbation, as it may cause instability of the compound soliton, see below. At the initial moment, Eq. (17) corresponds to soliton (13) with the HF component shifted by $\bar{x}(0)$. Note that, in its general form, ansatz (17) contains four free parameters, $A, B$ and $\bar{x}, k$, which provides the necessary balance with the set of four equations in system (14) – (16).

Substituting approximation (17) in moments-evolution equations (15) and (16), and assuming a small variation of the center-of-mass shift, $|\bar{x}| << 1/\sqrt{\lambda}$, we arrive, after simple manipulations, at the evolution equation for it,

$$\frac{d^2\bar{x}}{dt^2} - \frac{8}{15}\lambda B \bar{x} = 0.\quad(18a)$$

For a negative LF component (an LF *hole* with $B < 0$), it follows from Eq. (18a) that the HF component oscillates around the hole (which plays the role of a potential well) with frequency

$$\omega = 2\sqrt{-2\lambda B/15} \equiv 2\lambda\sqrt{2\gamma/(15\beta)}.\quad(18b)$$

For $B > 0$, which corresponds to a barrier-shaped LF component, Eq. (18a) gives rise to an exponentially growing solutions, implying that the soliton solution (13) is unstable.

## III. The system of coupled NLS-KdV equations

Under the commonly adopted assumption of the unidirectional wave propagation, the Bq equation may be reduced to one of the KdV type. In this case, the effective HF-LF interaction arises under the condition of the group synchronism of HF and LF waves, when the HF group velocity, $V_{\text{HF}} = (\partial \omega_{\text{HF}}/\partial k)_{k_0}$, is equal to the phase velocity of the LF, $V_{\text{LF}} = \omega_{\text{LF}}/k_{\text{LF}}$, running in the same direction: $V_{\text{HF}} = V_{\text{LF}} \equiv V_{\text{SIN}}$. This case takes place for the SW-IW interaction in stratified liquids [19] (an essential difference of our system from the one derived in Ref. [19] is we take into regard the self-interaction of the HF waves, accounted for by terms $\sim \alpha$ in Eq. (1) and in Eq. (19) written below; note that the soliton solutions given by Eqs. (13) and (20) for the NLS-Bq and NLS-KdV systems, respectively, do not exist for $\alpha = 0$). In the reference frame moving with the group-synchronism velocity $V_{\text{SIN}}$, $\xi = x - V_{\text{SIN}}t$, and neglecting the second time derivative, system (1) reduces to the following one, in which only unidirectional propagation is admitted for the LF waves [17]:

$$2i\frac{\partial U}{\partial t} - \frac{\partial^2 U}{\partial \xi^2} - 2\alpha|U|^2 U + 2nU = 0,\quad(19a)$$

$$2V_{\text{SIN}}\frac{\partial n}{\partial t} - 6\beta\frac{\partial(n^2)}{\partial \xi} + \gamma\frac{\partial^3 n}{\partial \xi^3} = -\varepsilon\frac{\partial(|U|^2)}{\partial \xi}.\quad(19b)$$

The effective nonlinearity of the LF waves is characterized by ratio $\gamma/\beta$ of coefficients in Eq. (19b). In terms of experimentally observable parameters for long IWs, their nonlinearity is determined by the Ursell number, $H\lambda^2/h^3$, where $H$ is the wave's amplitude, $\lambda$ the wavelength, and $h$ effective depth [23].

Group velocities of the coupled surface and internal waves may be in resonance at some wavelengths, $\Lambda_{\text{SW}}$ and $\Lambda_{\text{IW}}$. Taking a characteristic value for the Brunt-Väisälä (buoyancy) frequency, $\omega_{\text{BV}} \sim 0.01$ Hz, at the interface between the top mixed layer and the underlying undisturbed one in the ocean (at the depth of a few hundred meters) [21], and wavelength $\Lambda_{\mathrm{IW}} \sim 100$ m, one has $V_{\mathrm{IW}} \sim 15$ cm/s [20]. Equating this to the group velocity produced by the classical dispersion relation for the SW on deep water, $\omega_{\mathrm{SW}} = \sqrt{gk}$, we conclude that the corresponding characteristic frequency is $\omega_{\mathrm{SW}} \sim 1$ Hz, which exceeds the above-mentioned IW frequency, $\omega_{\mathrm{BV}}$, by two orders of magnitude, thus completely justifying the HF-LF distinction. The difference in the respective wavelengths is dramatic too, the estimate yielding $\Lambda_{\mathrm{SW}} \sim 1$ m. However, the disparities in the frequencies and wavelengths do not destroy the HF-LF coupling, once the group-velocity resonance holds.

## A. The soliton solution of the NLS-KdV system

For $\alpha \neq 0$ system (19) has a two-component soliton solution:

$$U = A\operatorname{sech}\left[\sqrt{\lambda}(\xi - Vt)\right]\exp(-i\lambda t - iV\xi), \quad n = B\operatorname{sech}^2\left[\sqrt{\lambda}(\xi - Vt)\right], \tag{20}$$

where $\lambda$ is a free parameter, similar to $\lambda$ in Eq. (13), $A \equiv \pm\sqrt{(\lambda/\alpha)(1-\gamma/\beta)}$, $B \equiv -\lambda\gamma/\beta$, $V \equiv -2\beta B/V_{\mathrm{SIN}} + \varepsilon A^2/(2BV_{\mathrm{SIN}})$. Solution (20) can be obtained from Eq. (13) for weak dispersion and nonlinearity of the LF waves, and weak HF-induced striction force acting on the LF component, when the velocity of soliton (13) is close to the velocity of the unidirectional linear LF waves, under condition $\left|4\gamma\lambda - (\varepsilon/\alpha)(\beta/\gamma - 1)\right| \ll V_{\mathrm{LF}}^2$.

## B. Dynamics of perturbed solitons with a relative shift between the HF and LF components of the NLS-KdV system

System (19), in which Eq. (19a) is rewritten in the Madelung form (cf. Eqs. (1c) and (1d)), with zero boundary conditions at infinity, $(n,U)\big|_{\xi\to\pm\infty} \to 0$, gives rise to the following integral relations for field moments, that are derived, similar to Eqs. (14)-(16), derived above for the NLS-Bq system, by means of the straightforward spatial integration:

$$\frac{d}{dt}\int_{-\infty}^{+\infty}|U|^2 d\xi = 0, \quad \frac{d}{dt}\int_{-\infty}^{+\infty} n\, d\xi = 0, \tag{21}$$

$$\frac{d}{dt}\int_{-\infty}^{+\infty}\left(\varepsilon K|U|^2 - V_{\mathrm{SIN}} n^2\right)d\xi = 0, \tag{22}$$

$$\frac{d}{dt}\int_{-\infty}^{+\infty} K|U|^2 d\xi = -\int_{-\infty}^{+\infty} n\frac{\partial\left(|U|^2\right)}{\partial\xi} d\xi, \tag{23}$$

$$\frac{d}{dt}\int_{-\infty}^{+\infty} \xi|U|^2 d\xi = -\int_{-\infty}^{+\infty} K|U|^2 d\xi. \tag{24}$$

In the adiabatic approximation, an approximate perturbed solution (*ansatz*) for the NLS-KdV may be looked for in form of sech-like pulses with a shift of the HF component with respect to the LF counterpart (as above), and temporal variation of widths and amplitudes of the two components:

$$U = a(t) A \operatorname{sech}\left[\frac{\sqrt{\lambda}(\xi - Vt - \bar{\xi}(t))}{\Delta(t)}\right] \exp[-i\lambda t - iV\xi + ik(t)\xi], \quad n = b(t) B \operatorname{sech}^2\left[\frac{\sqrt{\lambda}(\xi - Vt)}{\Delta(t)}\right], \quad (25)$$

where $\bar{\xi}(t) = N^{-1} \int_{-\infty}^{\infty} \xi |U|^2 d\xi$ is the center-of-mass coordinate of HF component, $k(t) = K + V$, $a^2(t)\Delta(t) \equiv \text{const}$, $b(t)\Delta(t) \equiv \text{const}$, $a(0) = b(0) = \Delta(0) = 1$. This is a generalization of the similar ansatz (17), adopted above for the analysis of the stability of the NLS-Bq solitons. At the initial moment, Eq. (25) corresponds to soliton (20) with the HF component shifted by $\bar{\xi}(0)$. Note that, in its general form, ansatz (25) may be used in the combination with the set of five equations in system (21) – (24), as it contains five free parameters: $a, b, \Delta$ and $\bar{\xi}, k$. Actually, most important parameters are, as in the case of the NLS-Bq system, center-of-mass shift $\bar{\xi}(t)$ and wavenumber $k(t)$, to which width $\Delta(t)$ is added. Substituting the ansatz in Eq. (22), we derive an integral of motion

$$a^2 = b = \frac{1}{\Delta} = 1 + \frac{3\varepsilon V_{\text{SIN}} A^2}{2B^2}(k - k_0), \quad (26)$$

where $k_0 = k(0)$. Further, the same ansatz (25) in Eqs. (23) and (24), and assuming a small variation of the center-of-mass shift, $|\bar{\xi}| \ll \Delta/\sqrt{\lambda}$, we eventually derive the dynamical equation,

$$d^2\bar{\xi}/dt^2 - (8/15)\lambda B \bar{\xi} = 0, \quad (27)$$

for the perturbed NLS-KdV soliton, cf. Eq. (18a).

It follows from Eq. (27) that, in the case of a negative LF component (an LF hole, $B < 0$, which plays the role of the potential well), the HF component oscillates around it with frequency

$$\omega = 2\sqrt{-2\lambda B/15} \equiv 2\lambda\sqrt{2\gamma/(15\beta)}, \quad (28)$$

which, incidentally, coincides with oscillation frequency (18b) obtained above for the perturbed NLS-Bq soliton. In particular, for initial pulse (25) with $k_0 \neq 0$ and $\bar{\xi}(0) = 0$, the temporal variation of the soliton's parameters is given by

$$\bar{\rho} = -\frac{k_0}{\omega}\sin(\omega t), \quad a^2 = b = 1 - \frac{3\varepsilon V_{\text{SIN}} A^2 k_0}{B^2}\sin^2\left(\frac{\omega}{2}t\right). \quad (29)$$

For $B > 0$, which corresponds to a potential barrier induced by the LF component for the HF one, Eq. (27) gives rise to an exponentially growing solution, which implies that soliton solution (20) is unstable.

## IV. Numerical results

### A. Single-soliton dynamics
*1. A relative velocity shift between the soliton's components*

We used the Maple software shell, based on the finite-difference scheme, to solve numerically the initial-value problem for the coupled NLS-KdV system (19), with $V_{SIN}=1$, $\alpha=1$, $\gamma=1/10$, $\varepsilon=1/10$ and input

$$U(\xi,0)=\sqrt{1-[1/(10\beta)]}\operatorname{sech}(\xi)\exp\left[-i(1/4-\beta/2)\xi+ik_0\xi\right],\ n(\xi,0)=-[1/(10\beta)]\operatorname{sech}^2\xi. \quad (30)$$

The simulations were performed with zero boundary conditions, viz., $U(\xi=\pm 200,t)=0$, $n(\xi=\pm 200,t)=0$. Time and spatial steps and $\Delta t=0.02$ and $\Delta x=0.025$ were sufficient to produce reliable results (this was checked by reproducing the same results with smaller $\Delta t$ and $\Delta x$). The typical relative accuracy of the numerical results is ~ $10^{-5}$. For $k_0=0$ initial pulse (30) corresponds to soliton (20) with $\lambda=1$, while wavenumber $k_0\neq 0$ introduces a perturbation in the form of the relative velocity between the HF and LF components. The simulations were performed for $k_0=1/4$ and different values of the LF nonlinearity coefficient $\beta$ in Eq. (19b). In Figs. 1(a) and (b), numerical results for HF and LF space-time distributions, $|U(\xi,t)|$ and $n(\xi,t)$, respectively, are displayed for $\beta=1/5$. In this case, the LF component is negative, having $B<0$, which corresponds to an effective potential well for the HF components, as said above). The simulations reveal periodic variations of the soliton parameters [HF and LF amplitudes, and the relative spatial shift of the HF component, cf. Eqs. (25) and (26)] with period $T\approx 12$, which is very close to the analytical value, $2\pi/\omega$, predicted by Eq. (28). The largest variations of the HF intensity, $a^2(t)$, and LF amplitude, $b(t)$, are close to 15% and agree with the analytical results given by Eq. (29). Figure 2 corresponds to $\beta=-1/5$, which gives rise the LF component with $B>0$, which, as said above, created a potential barrier, instead of a well, for the HF field. In this case, a small initial velocity shift between the components gives rise to an instability of soliton (20), corroborating the above-mentioned analytical prediction produced for this case by Eq. (27).

Simulations of the evolution of perturbed solitons in the NLS-Bq system give rise to similar results (not shown here in detail), which corroborate the stability of the solitons predicted by Eqs. (18) for $B<0$, and the instability for $B>0$.

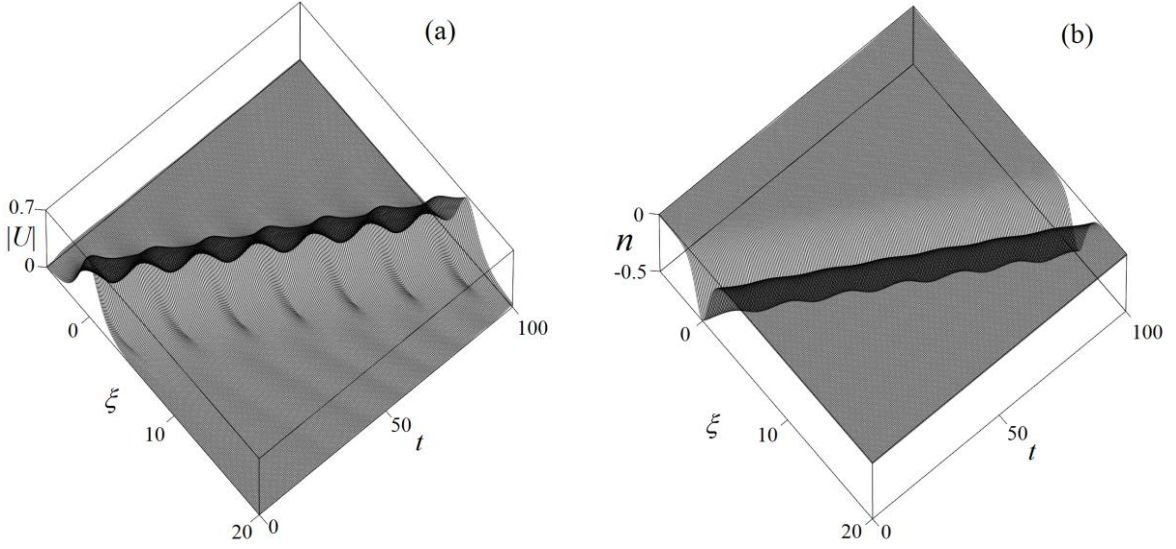

Fig.1. Results of simulations of the coupled NLS-KdV system (19) for spatio-temporal distributions of the HF field, $|U(\xi,t)|$ (a), and LF field, $n(\xi,t)$ (b), produced by initial pulse (30) with $\beta = 1/5$ (an LF hole). The perturbed evolution of the soliton is obviously stable in this case.

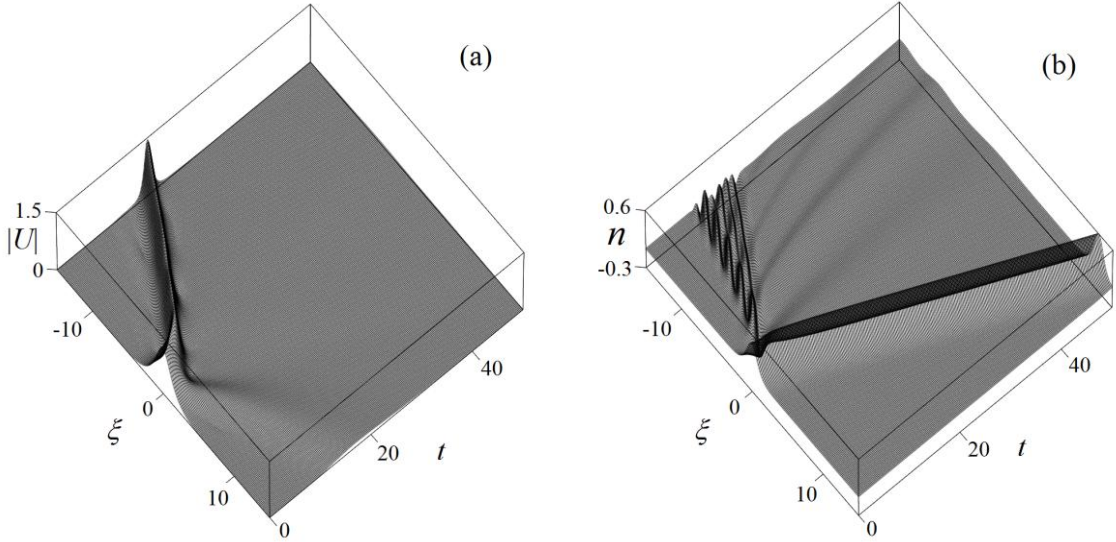

Fig.2. The same as in Fig. 1, but for $\beta = -1/5$ (an LF barrier), which gives rise to the instability of the perturbed soliton.

## 2. An amplitude perturbation of the soliton's components

Here we consider the evolution of the initially perturbed soliton taken as

$$U(\xi,0) = \sqrt{1/2}\,\text{sech}\,\xi \exp\left[-i(1/5)(D^2 - 1/4)\xi\right], \quad n(\xi,0) = -(D/2)\text{sech}^2\xi, \qquad (31)$$

in the framework of Eqs. (19) with $V_{\text{SIN}} = 1$, $\alpha = 1$, $\gamma = 1/10$, $\varepsilon = 1/10$, $\beta = 1/5$. For $D = 1$, pulse (31) corresponds to soliton (20) with $\lambda = 1$, while $D \neq 1$ introduces an amplitude perturbation.

For $D$ close to 1, the initial wave packet sheds off small-amplitude radiation wave packets and relaxes to a single soliton with parameters depending on $D$. For example, in Figs. 3(a) and (b) numerical results for the spatio-temporal distributions of $|U(\xi,t)|$ and $n(\xi,t)$, respectively, are displayed for initial conditions (31) with $D = 3/4$.

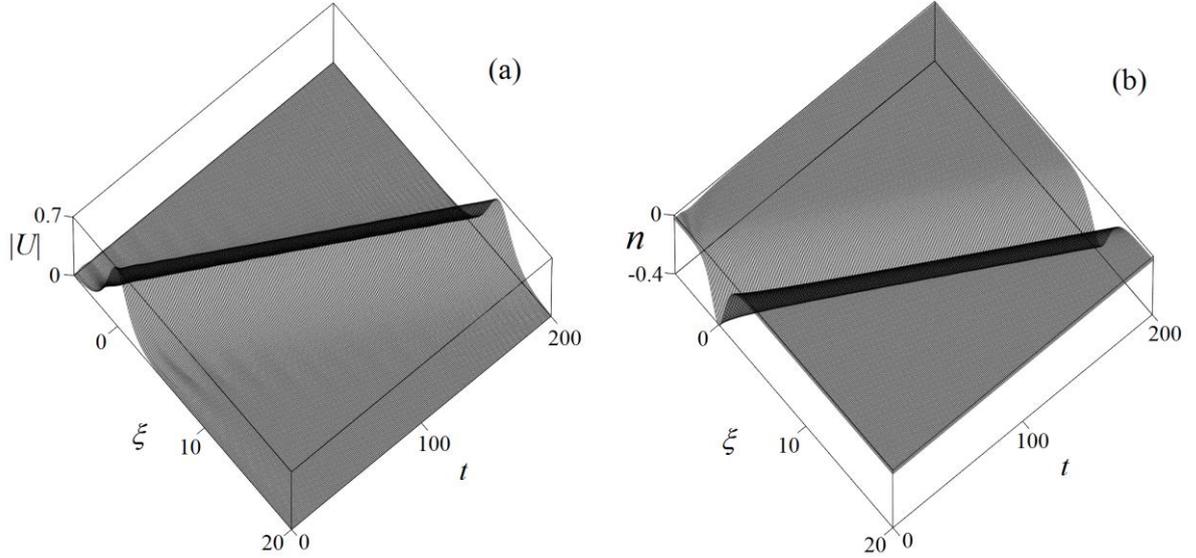

Fig. 3. Results of simulations of the coupled NLS-KdV system (19) for the spatio-temporal distributions of the HF and LF fields, $|U(\xi,t)|$ (a) and $n(\xi,t)$ (b), respectively, produced by initial pulse (31) with the amplitude perturbation corresponding to $D = 3/4$.

For interpretation of the numerical results with $D$ close to 1, we consider dynamics of the sech pulse taken as

$$|U| = a(t)A\text{sech}\left[\frac{\sqrt{\lambda}(\xi - \bar{\xi}(t))}{\Delta_U(t)}\right], \quad n = b(t)B\text{sech}^2\left[\frac{\sqrt{\lambda}(\xi - \bar{\xi}(t))}{\Delta_n(t)}\right], \qquad (32)$$

where $a(0) = 1$, $b(0) = D$, $\Delta_n(0) = \Delta_U(0) = 1$, and

$$a^2(t)\Delta_U(t) = a^2(0)\Delta_U(0) \equiv 1, \quad b(t)\Delta_n(t) = b(0)\Delta_n(0) \equiv D. \qquad (33)$$

Expression (32) at the initial moment corresponds to soliton (20) with the amplitude perturbation in the LF component, $|U| = A\text{sech}(\sqrt{\lambda}\xi)$, $n = DB\text{sech}^2(\sqrt{\lambda}\xi)$. Parameters of the stationary pulse, $a_*$, $(\Delta_U)_*$, $n_*$ and $(\Delta_n)_*$, into which the initial one will relax, can be estimated from the balance of the dispersion, nonlinearity and trapping-potential terms in Eq. (19a), and the balance of the dispersion and nonlinearity in Eq. (19b): $1/(\Delta_U^2)_* \approx \alpha a_*^2 A^2 - b_* B$, $\gamma/(\Delta_n^2)_* \approx -\beta b_* B$. Then, using relations $a_*^2(\Delta_U)_* = a^2(0)\Delta_U(0) \equiv 1$ and $b_*(\Delta_n)_* = b(0)\Delta_n(0) \equiv D$, see Eq. (33), we obtain the following values:

$$a_*^2 \approx 1 + \frac{D^2 - 1}{1 + \beta/\gamma}, \quad (\Delta_U)_* \approx 1/a_*^2, \quad b_* \approx D^2, \quad (\Delta_n)_* \approx 1/D. \qquad (34)$$

In particular, for initial pulse (31) with $D=3/4$ the stationary-pulse parameters predicted by analytical estimates (34) are: $a_* \approx 0.90$, $(\Delta_U)_* \approx 1.17$, $b_* \approx -0.56$, $(\Delta_n)_* \approx 1.33$. They are very close to the numerical found results: $a_{num} \approx 0.89$, $(\Delta_U)_{num} \approx 1.19$, $b_{num} \approx -0.60$, $(\Delta_n)_{num} \approx 1.31$.

For a strong-amplitude perturbation, with $D \gg 1$, the initial wave packet sheds off a large amount of radiation and splits into several quasi-stationary pulses with small intrinsic oscillations. This is made possible by the large amount of the LF field added to the input. For example, in Figs. 4(a) and (b) numerical results are displayed for the space-time distributions of $|U(\xi,t)|$ and $n(\xi,t)$, respectively, for input (31) with $D=3$. In this case, the initial wave packet relaxes into two quasi-stationary solitons with small oscillations.

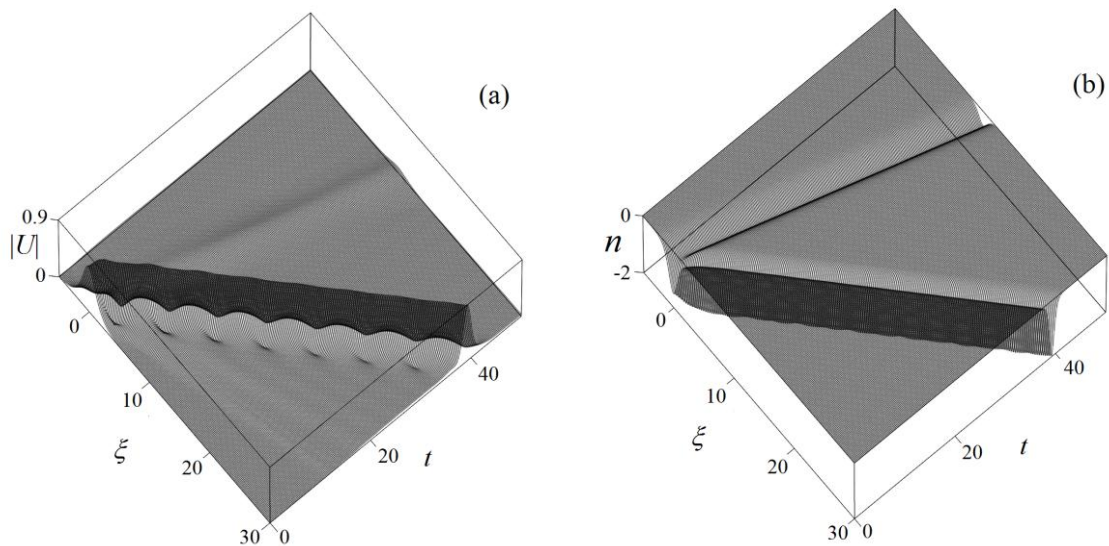

Fig. 4. The same as in Fig. 3, but for the strongly perturbed input (32) with $D=3$. In this case, it splits into two solitons which exhibit residual intrinsic vibrations.

Lastly, wave packet (31) with $D \ll 1$ relaxes into an HF soliton of the NLS type, in a combination with delocalized LF waves. Two-component solitons do not emerge in this case, as the LF component of the input is too weak

## B. Soliton-soliton interaction
### 1. The NLS-KdV system (19)

Interactions between stable solitons are an issue of obvious interest to physical realizations of the system. Here we consider collisions between solitons with initial distance $2\xi_0$ and phase shift $\varphi_0$ between them, and different values of $\lambda$ (hence, different velocities) in the framework of the NLS-KdV system (19) with $V_{SIN}=1$, $\alpha=1$, $\gamma=1/10$, $\varepsilon=1/10$, $\beta=1/5$. The respective input is

$$U(\xi,0)=\sqrt{\lambda_1/2}\operatorname{sech}[\sqrt{\lambda_1}(\xi-\xi_0)]\exp[-iV_1(\xi-\xi_0)+i\varphi_0]+\sqrt{\lambda_2/2}\operatorname{sech}[\sqrt{\lambda_2}(\xi+\xi_0)]\exp[-iV_2(\xi+\xi_0)],$$
$$n(\xi,0)=-(\lambda_1/2)\operatorname{sech}^2[\sqrt{\lambda_1}(\xi-\xi_0)]-(\lambda_2/2)\operatorname{sech}^2[\sqrt{\lambda_2}(\xi+\xi_0)], \quad V_{1,2}=\lambda_{1,2}/5-1/20. \qquad (35)$$

Results of the simulations are displayed in Figs. 5 and 6 in terms of spatio-temporal fields $|U(\xi,t)|$ (a) and $n(\xi,t)$ (b) for $\lambda_1 = 1$, $\lambda_2 = 5/4$, $\xi_0 = 5$, and different values of the phase shift: $\varphi_0 = 0$ (in-phase solitons) in Fig. 5, and $\varphi_0 = \pi$ (out-of-phase solitons) in Fig. 6. Naturally, the in-phase solitons attract each other, which eventually leads to their merger into one two-component pulse, with concomitant generation of an additional KdV soliton and nonsoliton radiation field in the LF component (obviously, the system admits solutions in which solely the LF field is different from zero). On the other hand, out-of-phase solitons interact quasi-elastically due to the mutual repulsion. The results shown in Figs. 5 and 6 also confirm that isolated solitons with the negative LF-field component, $n < 0$ (the ones observed prior to the collision), are stable.

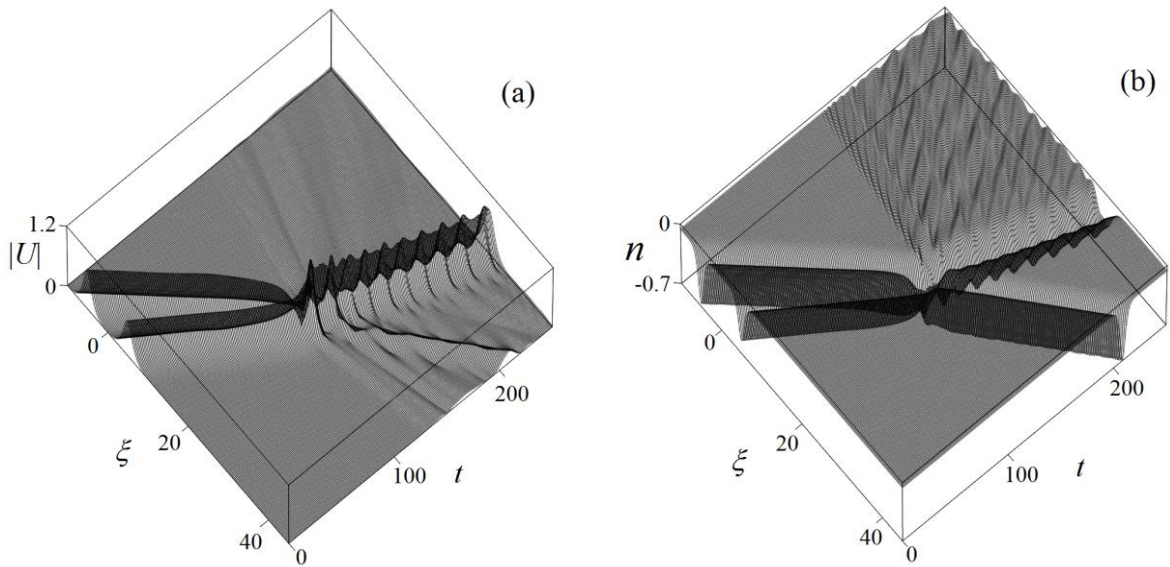

Fig.5. Panels (a) and (b) display spatio-temporal fields $|U(\xi,t)|$ (a) and $n(\xi,t)$ for a collision between in-phase solitons produced by simulations of Eqs. (19) with initial conditions (35), corresponding to $\varphi_0 = 0$. The solitons merge into a single two-component soliton and generate an additional LF (KdV) soliton, along with radiation field in the LF component.

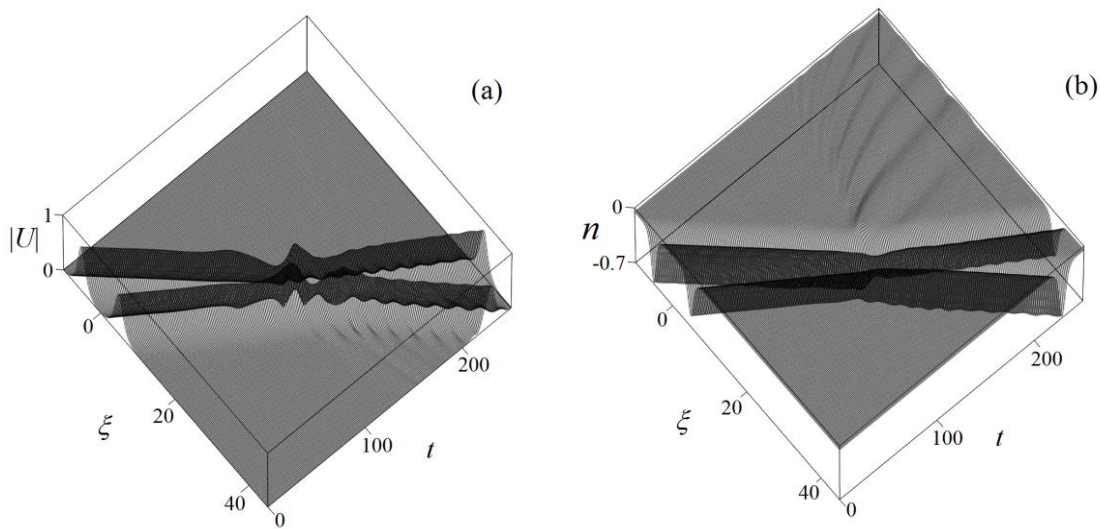

Fig.6. The same as in Fig. 5, but for a pair of out-of-phase solitons, with $\varphi_0 = \pi$. In this case, the collision is quasi-elastic.

## *2. The NLS-Bq system (1)*

The study of the soliton dynamics in the framework of the NLS-Bq system based on Eq. (1) in addition to the above results reported for the NLS-KdV system, is relevant, as it should help to check if the qualitative results are generic for the systems of the Zakharov's type, rather than being specific to a particular system. For this purpose, we here we consider soliton-soliton collisions with initial distance $2x_0$, phase $\varphi_0$ and different values of $\lambda$ (hence, different velocities) in the framework of the NLS-Bq system (1) with $V_{HF} = V_{LF} = 1$, $\alpha = 1$, $\gamma = -1/10$, $\varepsilon = 1/10$, $\beta = -1/5$. The respective input is

$$U(x,0) = \sqrt{\lambda_1/2}\,\text{sech}\left[\sqrt{\lambda_1}(x-x_0)\right]\exp\left[i(1-V_1)(x-x_0) + i\varphi_0\right]$$
$$+ \sqrt{\lambda_2/2}\,\text{sech}\left[\sqrt{\lambda_2}(x+x_0)\right]\exp\left[i(1-V_2)(x+x_0)\right],$$
$$n(x,0) = -(\lambda_1/2)\,\text{sech}^2\left[\sqrt{\lambda_1}(x-x_0)\right] - (\lambda_2/2)\,\text{sech}^2\left[\sqrt{\lambda_2}(x+x_0)\right],$$
$$\left.\frac{\partial n}{\partial t}\right|_{t=0} = \frac{V_1\lambda_1\sinh\left[\sqrt{\lambda_1}(x-x_0)\right]}{\cosh^3\left[\sqrt{\lambda_1}(x-x_0)\right]} + \frac{V_2\lambda_2\sinh\left[\sqrt{\lambda_2}(x+x_0)\right]}{\cosh^3\left[\sqrt{\lambda_2}(x+x_0)\right]}, \quad V_{1,2} = \sqrt{(9-4\lambda_{1,2})/10}. \quad (36)$$

Comparing this to its counterpart (35) for the NLS-KdV system, we notice that, in the present case, the input must include initial values of $\partial n/\partial t$, because the Bq component of the system, unlike the KdV-based one, is the evolution equation of the second order.

Results of the simulations are displayed in Figs. 7 and 8 in the reference system moving with the initial velocity of first soliton $V_1$, $\zeta = x - V_1 t$, in terms of spatio-temporal fields $|U(\zeta,t)|$ (a) and $n(\zeta,t)$ (b) for $\lambda_1 = 1$, $\lambda_2 = 0.8$ (i.e., a relatively small collision velocity), $x_0 = 5$, and different values of the phase shift between the colliding solitons: $\varphi_0 = 0$ (in-phase solitons) in Fig. 7, and $\varphi_0 = \pi$ (out-of-phase solitons) in Fig. 8. It is seen that, in the NLS-Bq system too, in-phase and out-of-phase collisions tend, respectively, to cause the merger or elastic rebound of the colliding solitons. The difference of the merger observed in Fig. 7 from the analogous outcome displayed for the NLS-KdV system in Fig. 5 is a smaller amount of radiation emitted in the course of the merger. Similarly, the elastic rebound observed in Fig. 8 (at a large distance between the colliding solitons, because the collision velocity is small), shows virtually no emission of radiation.

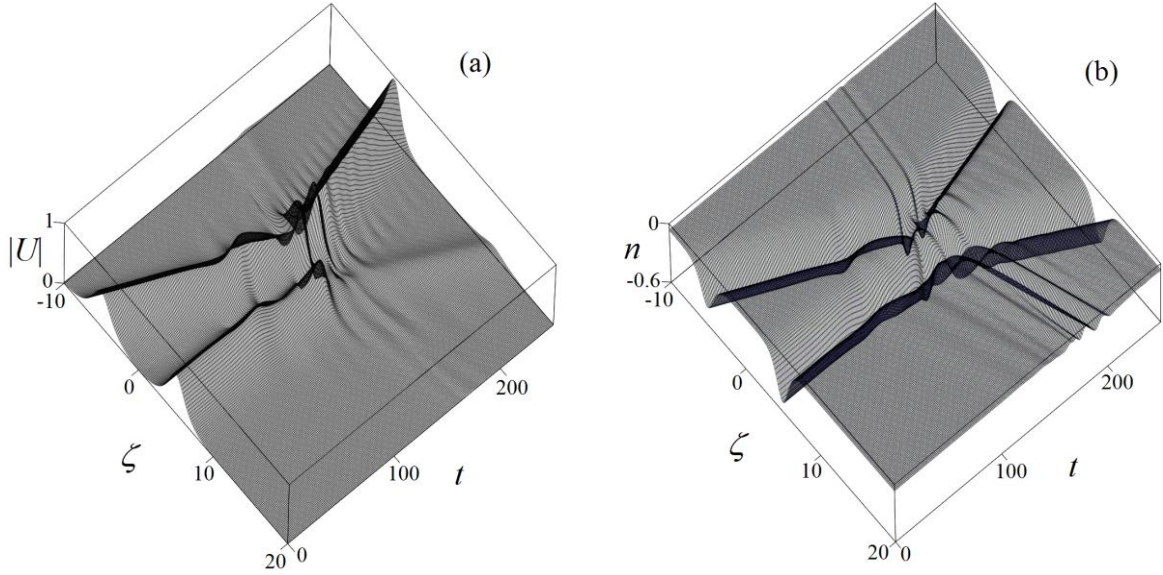

Fig.7. Spatio-temporal fields $|U(\zeta,t)|$ (a) and $n(\zeta,t)$ (b), produced by simulations of a collision of two in-phase solitons in the framework of Eq. (1) with initial conditions (36), corresponding to $\varphi_0 = 0$, and displayed in the reference frame moving along with the first soliton ($\zeta \equiv x - V_1 t$, see the text). They merge into a single two-component soliton and generate an additional LF (Bq) soliton, along with radiation field in the LF component.

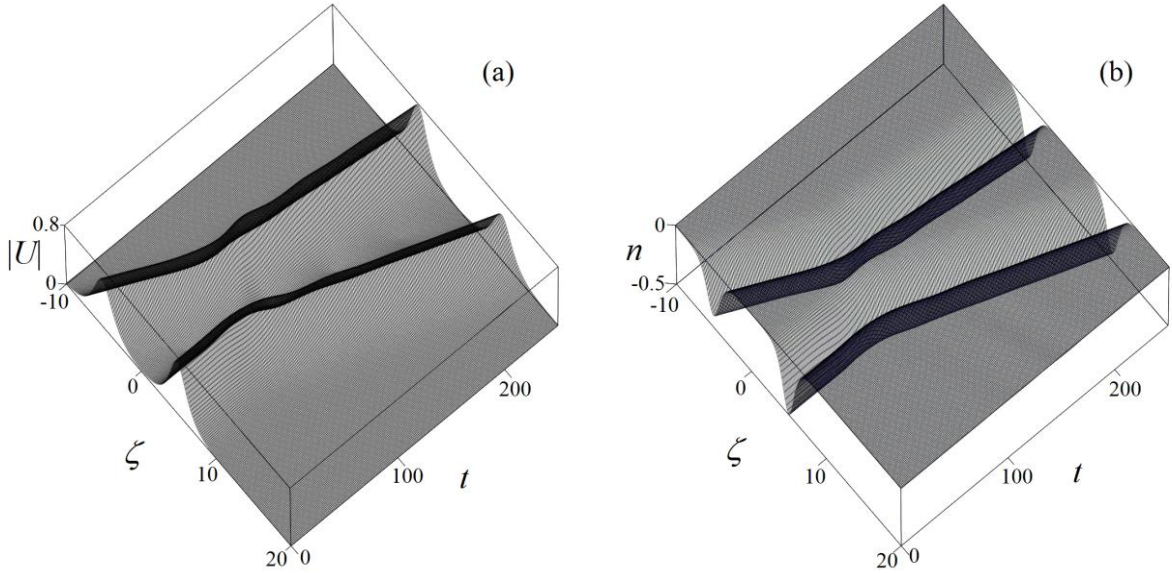

Fig.8. The same as in Fig. 7, but for a pair of out-of-phase solitons, with $\varphi_0 = \pi$. In this case, the collision is quasi-elastic.

## V. Conclusion

We have considered the interaction of intense HF (high-frequency) and LF (low-frequency) waves in the framework of the Zakharov-type system of coupled equations for the HF and LF fields. Such systems are relevant interaction models when the full underlying system, such as hydrodynamic equations for stratified liquids, admit a group- and phase-velocity resonance between the HF and LF wave modes. In the framework of the Zakharov's system, the HF field is governed by the NLSE with the self-phase modulation term and the potential term which accounts for the coupling to the LF field. The LF component is governed by the Bq (Boussinesq) or KdV equation, severally for the bi-directional or unidirectional propagation of the LF waves, with the quadratic term accounting for the striction-induced coupling to the HF field. New exact one-parameter families of two-components soliton solutions for the coupled HF and LF waves have been produced, as two-component generalizations of the one-component soliton solutions of the Bq and KdV equations. They do not represent the most generic family of solitons admitted by the NLS-Bq and NLS-KdV systems, as a generic family plausibly has two free parameters, rather than one (amplitude and velocity), and can be found only in a numerical form. However, the solutions represent the most general soliton sets which may be found in an analytical form, being relevant for this reason. The stability and perturbed dynamics of the solitons was investigated in detail, analytically (by means of the method of moments) and numerically. In particular, the solitons with the negative LF component, which plays the role of the potential well for the HF mate, are stable against the critical perturbation, which is spontaneous separation of the two components. On the other hand, the solitons are unstable if the LF component is positive, as in that case it acts on the HF field as a potential barrier. Perturbed evolution of the solitons initiated by a change of the amplitude of the LF component was analyzed in detail too. Collisions between solitons were studied by means of direct simulations, demonstrating a trend to the merger of in-phase solitons, and quasi-elastic interactions between out-of-phase ones.

As an extension of this work, it may be interesting, as mentioned above, to look for still more general two-component solitary-wave modes in the NLS-Bq and NLS-KdV systems, without confining the consideration to exact solutions, but rather using full numerical methods.

## Acknowledgements

This work was performed, in a part, in the framework of the Academic Fund Program at the National Research University – Higher School of Economics (HSE) in 2016-2017 (grant № 16-01-0002), and partly supported by a subsidy granted to the HSE by the Government of the Russian Federation for the implementation of the Global Competitiveness Program.